\RequirePackage{xcolor}

\documentclass[journal,twoside,web]{ieeecolor}
\usepackage{tuffc}
\usepackage{cite}
\usepackage{amsmath,amssymb,amsfonts}
\usepackage{algorithmic}
\usepackage{graphicx}
\usepackage{textcomp}

\usepackage{booktabs} 
\usepackage{fontawesome}
\usepackage{multirow}
\usepackage{float}

\def\BibTeX{{\rm B\kern-.05em{\sc i\kern-.025em b}\kern-.08em
    T\kern-.1667em\lower.7ex\hbox{E}\kern-.125emX}}
\begin{document}
\title{Self-Supervised Learning with Limited Labeled Data for Prostate Cancer Detection in High Frequency Ultrasound}
\author{Paul F. R. Wilson\text{\Large*}, Mahdi Gilany\text{\Large*}, Amoon Jamzad, Fahimeh Fooladgar, Minh Nguyen Nhat To, Brian~Wodlinger, Purang Abolmaesumi\textsuperscript{\dag}, Parvin Mousavi\textsuperscript{\dag}
\thanks{Manuscript submitted for review on September 30, 2022. This work was sponsored in part by the Natural Science and Engineering Research Council of Canada (NSERC), and in part by the Canadian Institutes of Health Research (CIHR).  }
\thanks{Paul Wilson, Mahdi Gilany, Amoon Jamzad and Parvin Mousavi are with the School of Computing, Queen's University, Kingston ON K7L 3N6 Canada (e-mail 1pfrw@queensu.ca, mousavi@queensu.ca).}
\thanks{Fahimeh Fooladgar, Minh Nguyen Nhat To, and Purang Abolmaesumi are with the Dept. of Electrical and Computer Engineering, University of British Columbia, Vancouver BC V6T 1Z4, Canada.}
\thanks{Brian Wodlinger is with Exact Imaging, Markham ON L3R 2N2, Canada.}
\thanks{\text{\large*} Equal contribution. The listing order is random.}
\thanks{\textsuperscript{\dag} Purang Abolmaesumi and Parvin Mousavi are co-senior authors.}}

\maketitle

\begin{abstract}
Deep learning-based analysis of high-frequency, high-resolution micro-ultrasound data shows great promise for prostate cancer detection. Previous approaches to analysis of ultrasound data largely follow a \emph{supervised} learning paradigm. Ground truth labels for ultrasound images used for training deep networks often include coarse annotations generated from the histopathological analysis of tissue samples obtained via biopsy. This creates inherent limitations on the availability and quality of labeled data, posing major challenges to the success of supervised learning methods. On the other hand, unlabeled prostate ultrasound data are more abundant. In this work, we successfully apply self-supervised representation learning to micro-ultrasound data. Using ultrasound data from 1028 biopsy cores of 391 subjects obtained in two clinical centres, we demonstrate that feature representations learnt with this method can be used to classify cancer from non-cancer tissue, obtaining an AUROC score of 91\% on an independent test set. To the best of our knowledge, this is the first successful end-to-end self-supervised learning approach for prostate cancer detection using ultrasound data. Our method outperforms baseline supervised learning approaches, generalizes well between different data centers, and scale well in performance as more unlabeled data are added, making it a promising approach for future research using large volumes of unlabeled data. 
\end{abstract}


\section{Introduction}
\label{sec:introduction}
\IEEEPARstart{P}{rostate cancer} (PCa) is the second most common cancer diagnosed in men worldwide~\cite{smith2018canadian}. Early and accurate detection and staging of PCa is critical to guide treatment decisions. The standard of care for diagnosing PCa is histopathological analysis of tissue samples using the Gleason grading system where microscopic patterns of the tissue are used to determine cancer grades (1 to 5), and the grades of the two dominant tissue patterns are added and reported as the Gleason score. 
Tissue samples are obtained via needle biopsy, typically under the guidance of transrectal ultrasound (TRUS). TRUS is used for navigation but not for targeting the biopsy, as historically it has lacked sufficient accuracy in identifying cancerous lesions~\cite{smeenge2012current}. Instead, freehand prostate biopsy is primarily \emph{systematic} where a number of biopsy cores are collected, in a specific pattern, from the prostate with the aim of obtaining enough samples to identify cancer, should it be present. Still, cancer is frequently missed and many patients with a negative biopsy will eventually require re-biopsy and be diagnosed with cancer \cite{wolters2010false,ahmed2017diagnostic}. In other cases, men undergo unnecessary biopsies for benign pathologies or indolent cancers where a watch-and-wait approach may be preferable. Biopsies carry the risk of adverse events~\cite{ahmed2017diagnostic}. Any improvements in the ability to detect or rule out cancer by direct analysis of ultrasound images would have a major impact on patient outcomes. 

A substantial body of literature has established the limitations of B-mode ultrasound for PCa detection~\cite{smeenge2012current}. On the other hand, analysis of raw radio-frequency (RF) echo data is more promising, as the frequency and phase information contained therein has been shown to correlate with tissue properties. This reasoning is the foundation for quantitative ultrasound (QUS) methods which compute envelope statistics and backscatter coefficients from RF data and associate them with tissue microstructure~\cite{oelze2012quantitative,oelze2016review}. QUS has shown to improve PCa detection compared to analysis of B-mode images~\cite{feleppa2009prostate}. In particular, manual features selected from QUS combined with machine learning methods has seen considerable success~\cite{feleppa2009prostate}. Although handcrafted feature selection allows for better explainability and suffers less from overfitting, it is restricted to a relatively small number of features and may miss unknown properties in the raw RF data that correlate with PCa. Deep learning approaches, on the other hand,  allow feature extraction from raw data and are increasingly outperforming classical models in other areas of medical imaging~\cite{litjens2017survey}. 

Other ultrasound-based methods, than QUS, have also been proposed to improve PCa detection. 
Doppler ultrasound has been used for cancer detection by measuring angiogenesis associated with tumour formation~\cite{kelly1993prostate,nelson2007targeted,khanduri2017evaluation}, while elastography-based methods have shown promise for detecting PCa by measuring tissue stiffness~\cite{pallwein2007real,salomon2008evaluation,aleef2022quasi}. Temporal enhanced ultrasound has also been applied for PCa detection by enhancing the resolution of imaging of tissue microsctructures \cite{moradi2008augmenting,azizi2017detection, azizi2018deep}. However, these methods all use conventional clinical imaging frequencies (9-14 MHz) that only afford limited spatial resolution, and may hinder the ability to robustly identify PCa. 

The emerging state-of-the-art is to combine TRUS with multiparametric MRI (mp-MRI), which has a higher sensitivity (88-96\% compared to 42-55\% for conventional TRUS) in detection of PCa~\cite{ahmed2017diagnostic}. Fusion of mp-MRI with TRUS enables biopsy targeting by identifying suspicious lesions in the prostate \cite{rai2021magnetic,siddiqui2013magnetic}. However, the availability of MRI is limited, and fusion biopsy requires image registration that can be prone to errors due to patient movement. The improvement of biopsy targeting using TRUS directly is therefore highly desirable.  

Recently developed high frequency ``micro-ultrasound" technology allows imaging the prostate at a much finer spatial resolution~\cite{klotz2020can}. Clinical studies show that micro-ultrasound has a sensitivity comparable to mp-MRI using the qualitative ultrasound based ``PRI-MUS" scoring system~\cite{abouassaly2020impact,eure2019comparison,ghai2016assessing}, and a recent meta-analysis of 13 published studies with 1125 participants concludes that micro-ultrasound guided biopsy has similar PCa detection rates as mp-MRI fusion biopsy~\cite{sountoulides2021micro}. However, analysis of RF data remain relatively unexplored for micro-ultrasound: these are limited to a single study using machine learning with QUS (Rohrbach et al.~\cite{rohrbach2018high}) and two studies (Shao et al.~\cite{shao2020improving}, and Gilany et al.~\cite{gilany2022towards}) using deep learning. While promising, there are key challenges to the development of machine-learning based cancer detection models that are clinically useful. We argue that not all of these challenges have been adequately addressed for micro-ultrasound, and present a self-supervised learning (SSL) approach as a solution. The challenges are as follows:



\paragraph{Weak Labeling}


Machine learning methods for PCa detection rely 
pathology annotations as ground truth labels for corresponding ultrasound data. These annotations are only coarse approximations of the distribution of cancer in the tissue. When using the label of "malignant", for instance, it is not known precisely which areas within the needle region were cancer and which were benign. This weak labelling can severely impact the robustness of deep learning models which tend to memorize incorrect labels~\cite{han2018co}.


\paragraph{Heterogeneity}
Prostate tissue includes normal tissue, benign conditions, precancerous changes, and cancers ranging from indolent to highly aggressive. Within each of these categories there is significant variability in tissue characteristics as well. Ultrasound is subject to noise and imaging artefacts that further increase heterogeneity in tissue appearance. It is, therefore, challenging to train models which generalize to unseen data, specifically to tissue variations that may appear as out-of-distribution (OOD). 

\paragraph{Distribution Shift}
Ultrasound data are prone to major distribution shifts due to differences in equipment, clinical settings, and patient populations\cite{blaivas2022deep}. A typical example is the distribution shift between clinical centers. Standard deep learning approaches are not robust to distribution shifts~\cite{recht2018cifar}, limiting the clinical translation of these models. 

\paragraph{Data Scarcity}
Obtaining labeled ultrasound data requires a biopsy and additional time-intensive annotation from a human expert. This strongly limits the availability of large datasets to train and evaluate models. Deep learning methods have historically relied on copious amounts of data to learn and generalize well; labeled-data scarcity is a major challenge on its own that also exacerbates other previously mentioned challenges. 

Previously various solutions to these challenges have been proposed. For weak labeling, Zou et al.~\cite{zou2021robust} propose a noisy annotation tolerant network for breast ultrasound segmentation. Javadi et al. propose to use multi-instance learning~\cite{javadi2020multiple} and co-teaching~\cite{javadi2021training} for PCa detection. Uncertainty estimation allows a model to express uncertainty when seeing OOD data rather than making false predictions; such methods have been applied to prostate~\cite{fooladgar2022uncertainty,gilany2022towards} and breast~\cite{mojabi2020tissue,zhang2021tumor} ultrasound to address heterogenity. Shao et al.~\cite{shao2020improving} propose to handle inter-center distribution shift on micro-ultrasound data by training an adversarial auxiliary model to predict data center origin from the hidden features of a cancer detection model, thereby encouraging the detection model to be invariant under distribution shifts. For data scarcity, transfer learning from larger natural image datasets~\cite{ayana2021transfer} or learning from synthetic data~\cite{behboodi2019ultrasound} has been applied to B-mode (but not RF) ultrasound. While these strategies have addressed individual angles of the problem, none are a unified solution, and none adequately address data scarcity. 

Additionally, these approaches to PCa detection all follow a \emph{supervised learning} (SL) approach, while \emph{self-supervised} methods remain unexplored. However, there are intuitive, theoretical, and empirical reasons that make self-supervised learning a potentially unified solution. SSL allows learning without labels, sidestepping issues of weak label memorization. SSL has been empirically found to improve robustness to label corruptions~\cite{hendrycks2019using}, and improve model uncertainty. SSL has also been shown to be more robust to dataset imbalance~\cite{liu2021self} which may improve performance on under-represented natural tissue variations. Self-supervised models can be more robust to dataset-level distribution shift~\cite{zhong2022self} and have better transfer learning performance~\cite{ericsson2021well} than their supervised counterparts. The benefits of transfer learning using SSL on domain-specific data have been shown for a variety of x-ray and histology slide image tasks~\cite{azizi2022robust}. Finally, and possibly the most compelling, is that SSL enables learning with much more abundant unlabeled data, addressing the data scarcity challenge directly.

\subsection{Contributions}

We conduct a study on SSL methods for RF micro-ultrasound data. To the best of our knowledge, this is the first application of SSL for automatic PCa detection using RF ultrasound (either at micro-ultrasound \emph{or} conventional frequencies). Through extensive experiments on data from two clinical centers involving 391 total subjects (1028 total biopsy cores), we demonstrate that:

\begin{itemize}
    \item 
    SSL significantly improves PCa detection compared to supervised learning (SL) alone. By using unlabeled data, SSL allows the model to learn features from a greater volume of data and wider range of tissue types, addressing the issues of data scarcity and heterogeneity. 
    \item
    Even when using matched amounts of data, SSL still significantly outperforms SL by avoiding label memorization and reducing the impact of weak labels. 
    \item 
    SSL models outperform SL models when used for transfer learning between datasets. By capturing useful general features of RF ultrasound data that are independent of specific distributions, SSL alleviates distribution shift.

\end{itemize}

\section{Materials}

\subsection{Data Acquisition}

We use data from a multi-center clinical trial (Multi-Center Trial of High-resolution Transrectal Ultrasound Versus Standard Low-resolution Transrectal Ultrasound for the Identification of Clinically Significant Prostate Cancer, {clinicaltrials.gov}, NCT02079025). Data collected from 391 patients at two sites were included in our study: Urology of Virginia, Virginia Beach, USA (UVA) and Centre de Recherche sur le Cancer, Quebec City, Canada (CRCEO). Subjects underwent systematic TRUS-guided prostate biopsy using the ExactVu micro-ultrasound system (ExactVu, Markham, Canada). The system consists of a side-mounted linear array with 512 evenly-spaced transducers covering an area of 4.6~cm. The system operates with a pulse frequency up to 29~MHz (compared to the standard 6-9~MHz range of conventional ultrasound), capturing a high-resolution ultrasound image of the prostate. For each biopsy location, raw Radio Frequency (RF) ultrasound images of the tissue were saved immediately prior to the biopsy gun being fired. The biopsy needle enters the tissue at a fixed angle relative to the imaging plane. The approximate needle trace region was determined using this known angle and the penetration depth. RF scans consist of 512 RF lines (lateral dimension) with 10016 samples (axial dimension). Therefore, each RF scan is an image of shape $(512, 10016)$ consisting of echo intensity values. These images correspond to a physical tissue extent of 46~mm laterally and 28~mm axially. 

 Patients underwent a standard 12-core biopsy procedure. Cores were analyzed histopathologically to determine the Gleason score, primary and secondary Gleason grades, and approximate percentage of cancer (termed the ``involvement"). Additional data available for each patient are the prostate specific antigen (PSA) test score, age of the patient, and indication of family history of cancer. Among the $3804$ total cores obtained, the vast majority ($86.7\%$) are non-cancerous. Cancerous cores with Gleason scores 7, 8, 9 and 10, respectively, make up $10\%, 2\%, 0.9\%$ and $0.1\%$ of the remaining cores. Data are stratified into training and testing sets by patient rather than by core to ensure there is no data leakage between sets. This selection is done by randomly choosing patients for the test set until the number of cancer cores in the test set compared to the remaining cores reaches the desired ratio of approximately $25\%$ to $75\%$. To balance the cancerous and benign classes, we under-sample the benign cores to match the number of cancerous cores.   

\subsection{Data Preprocessing}

\begin{figure}
    \centerline{
    \includegraphics[scale=0.3]{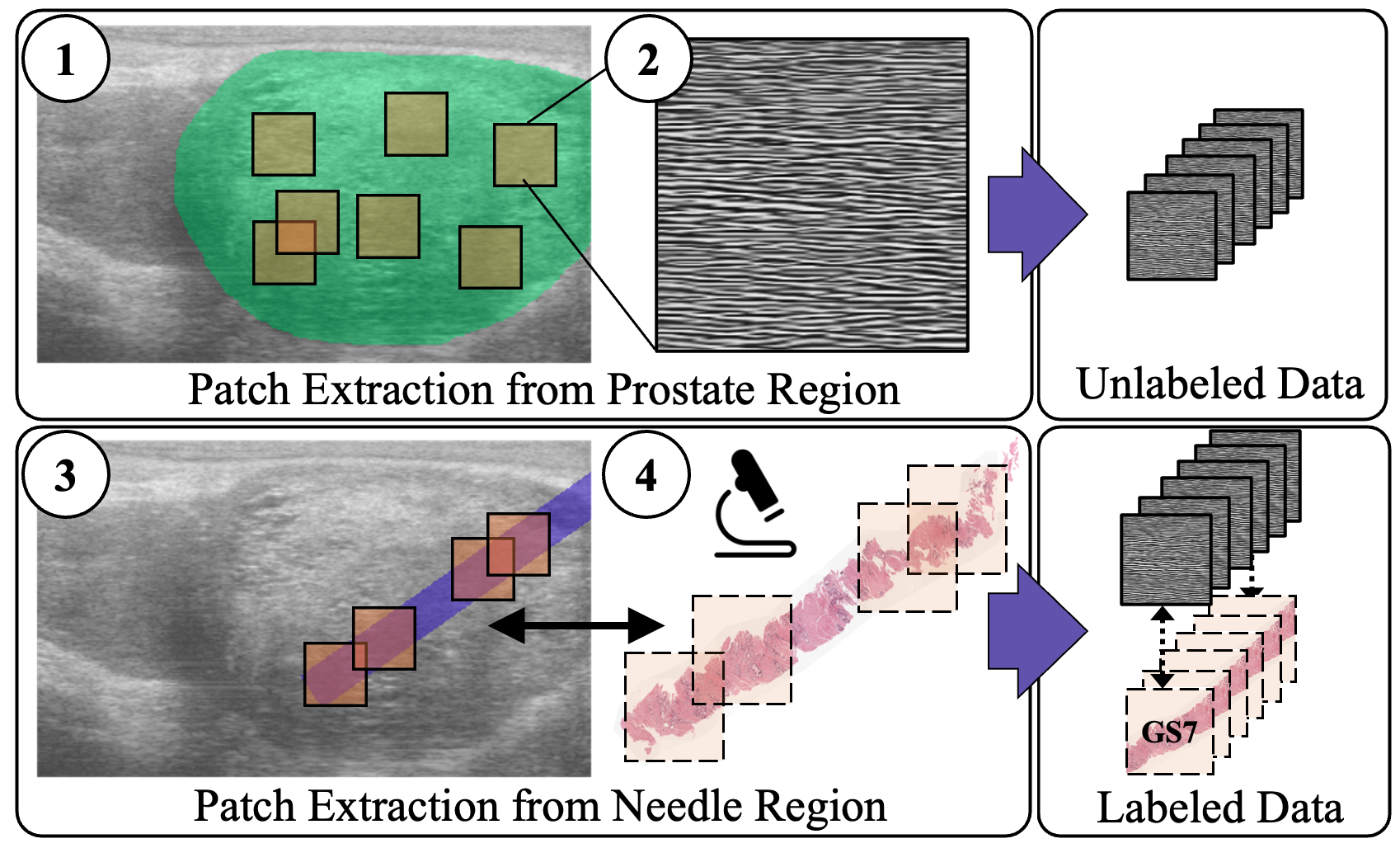}}
    \caption{Summary of data preparation from raw RF ultrasound to model input: (1) For unlabeled data, patches are selected from the prostate region; (2) close up of an RF ultrasound patch; (3) for labeled data, patches are selected from the needle trace region and paired with (4) tissue labels from histopathology of the corresponding tissue sample.}
    \label{fig:my_label}
\end{figure}

Patches corresponding to a tissue area of $5~\text{mm} \times 5~\text{mm}$ are extracted from the RF images as input to our models. A prostate segmentation mask was manually drawn for each core. The prostate occupied an average $45.46\%$ of the area of each image, while the needle region occupied $7.2\%$. For the unlabeled dataset, patches were extracted from anywhere within the prostate region. For the labeled dataset, patches were selected from within the intersection needle trace region and prostate region, and labeled 0 (benign) or 1 (malignant) based on the pathology findings of the core. We considered a patch to be within the needle region if it overlaps by at least $66\%$ with the needle trace mask. The labeled and unlabeled datasets corresponding to the UVA and CRCEO centers are denoted by $\mathcal{D}^{\text{UVA}}_{\text{needle}},$ $\mathcal{D}^{\text{CRCEO}}_{\text{needle}},$ $\mathcal{D}^{\text{UVA}}_{\text{prostate}},$ $\mathcal{D}^{\text{CRCEO}}_{\text{prostate}},$ respectively. 

Patches were uniformly reshaped to a size of $256 \times 256$ pixels using linear interpolation. Note the RF lines are generally band-limited, so no loss of information due to aliasing occurs during this resizing. Each patch was instance-normalized by computing its mean and standard deviation, truncating pixels which fall above or below four standard deviations from the mean, and rescaling to the range $(0, 1)$.

\subsection{Data Augmentations}\label{sec:data_augs}

Selection of data augmentations is an important consideration when using SSL methods for computer vision as they drive the learning objective: Networks are trained to extract features which are invariant under different augmentations. In doing so the network learns to extract high-level semantic features that do not depend on the specific pixel-level features of an image. Augmentations should significantly distort the input image such that the task is difficult, but not so much as to destroy high-level features relevant to downstream tasks.

We reasoned that the standard natural image augmentation pipeline for SSL (resized cropping, random application of Gaussian filters and random color jitter~\cite{chen2020simple, bardes2021vicreg}) is unlikely to be the correct choice for RF ultrasound, as these transformations alter frequency content which contains important tissue information. Instead, we use a combination of rigid transformations and masking:
\begin{itemize}
    \item \verb|random_translation|: Translating the image by a factor of up to 0.2 in the horizontal and vertical directions independently, and filling the empty pixel values with 0.5
    \item 
    \verb|random_erasing|: Selecting and filling with the value 0.5 a rectangular patch with height and width a factor between 0.02 and 0.1 the image size
    \item 
    \verb|random_vertical_flip|
    \item \verb|random_horizontal_flip|
\end{itemize}

We also introduce several augmentations handcrafted specifically to the physics of RF ultrasound. Ultrasound physics-inspired augmentations have been proposed previously for B-mode~\cite{ultrasound_physics} but not RF. Our approach considers the decomposition of an RF line into envelope and instantaneous frequency. Let $s(t)$ for $t \in \mathbb{R}$ denote an RF echo (the following analysis can be easily converted to discrete time samples, but for clarity we use the continuous time variable). Then the analytic representation of $s$, denoted by $s_a(t)$, is given by

\begin{equation}
    s_a(t) = s(t) + j \mathcal{H}(s)(t), 
\end{equation}
where $\mathcal{H}(\cdot)$ is the Hilbert transform operator \begin{equation}
    \mathcal{H}(f)(t) = \frac{1}{\pi} \text{p.v.} \int_{-\infty}^{\infty} \frac{f(t)}{\tau - t} d\tau, 
\end{equation}
(where $p.v$ denotes the Cauchy principle value definition of the integral), and $j$ is the imaginary unit. When $s$ is real, so is $\mathcal{H}(s)$; thus the original signal can be recovered via $s(t) = \mathfrak{R} s_a(t)$. By writing the analytic representation as $s_a(t) = A(t)e^{j\phi(t)}$, we can call $\phi(t)$ the \emph{instantaneous phase}, $\phi'(t)$ the \emph{instantaneous frequency}, and $A(t)$ the \emph{instantaneous amplitude} (or envelope) of the signal. 
Our augmentations are: 
\begin{itemize}
    \item \verb|random_phase_shift| -- Shifts the phase of the signal without changing the instantaneous envelope or frequency:
    \begin{equation*}
        T(s)(t) = \mathfrak{Re}\big( s_a(t) e^{i\theta} \big); \theta \sim \text{Uniform}(0, 2\pi)
    \end{equation*}
    
    \item \verb|random_envelope_distort| -- Alters the envelope but not the phase or frequency (``Noise" is low-pass filtered white noise): 
    \begin{equation*}
        T(s)(t) = \mathfrak{Re} \big( s_a(t) (1 + n(t))\big); n \sim \text{Noise}
    \end{equation*}

\end{itemize}
    
Note that when applied to patches, the same randomly sampled transformation is applied to each RF line in the patch. 

All of these augmentations are used together during self-supervised and (optionally) supervised training: to perform the augmentation, for each category, either that category is skipped with probability 0.5 or a transformation is randomly chosen and applied from that category.

\section{Methods}

\begin{figure}
    \centering
    \includegraphics[scale=0.38]{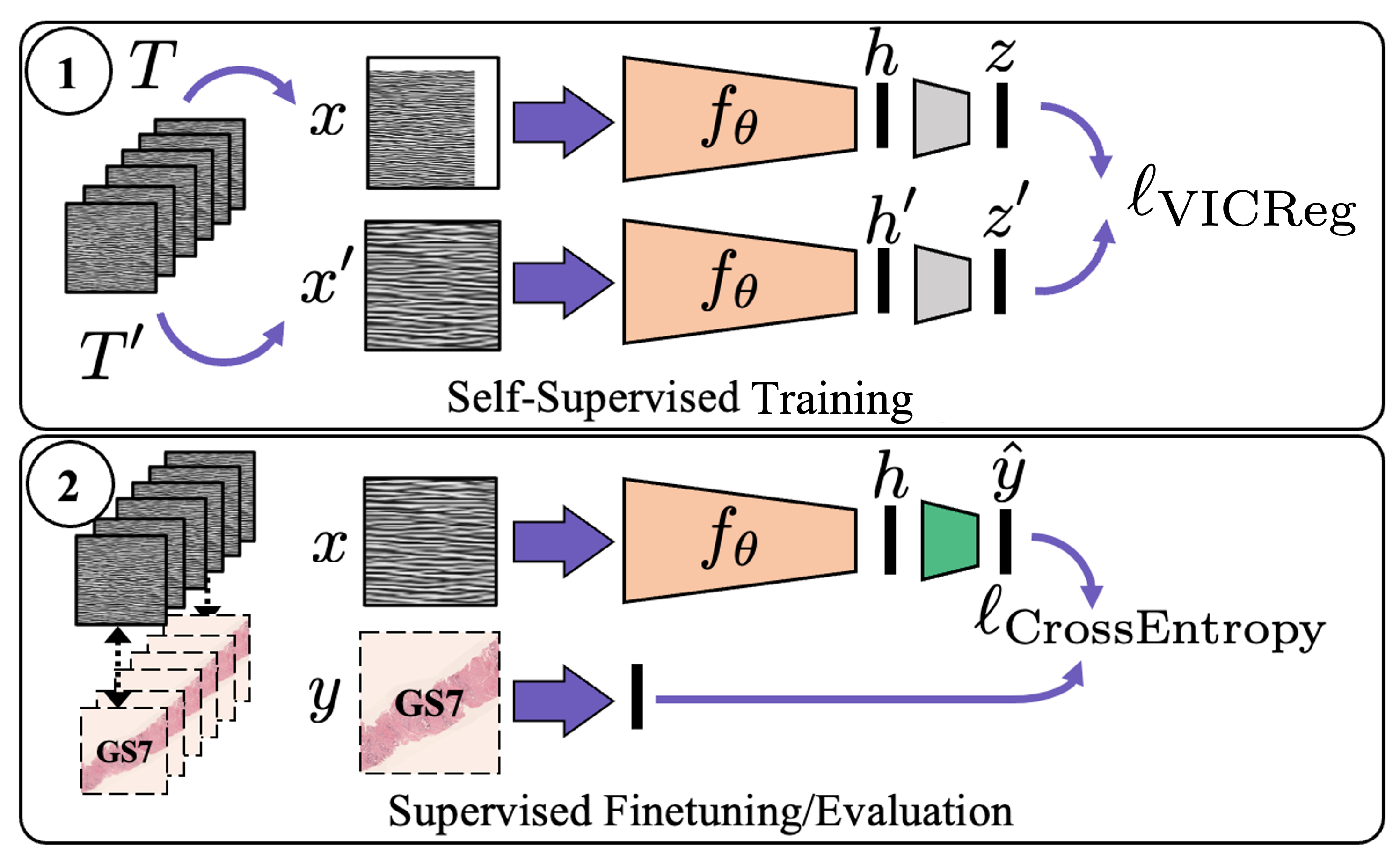}
    \caption{Summary of our learning approach. (1) Self-supervised learning: a feature extractor is trained to reduce the VICReg loss~\cite{bardes2021vicreg} between representations from pairs of augmented views of RF ultrasound data. (2) Supervised fine-tuning/evaluation: the model is trained to minimize the cross-entropy loss between true histopathology labels and predicted labels, using the (possibly frozen) pretrained feature extractor, $f_\theta$, as a backbone network.}
    \label{fig:method}
\end{figure}

We use SSL to learn high quality feature representations for RF ultrasound using unlabeled data. This allows us to sidestep any problems associated with weak labels while drastically increasing the pool of data available for training. Following standard practice in SSL, we use a two-stage pipeline consisting of (1) self-supervised training (also referred to as ``pretraining"), then (2) supervised fine-tuning and evaluation. This pipeline is illustrated in Figure \ref{fig:method}.


\subsection{Self-Supervised Representation Learning}

We use SSL to train a backbone network (in our case, a modified ResNet~\cite{he2016deep} architecture with $\sim$~6M parameters) to extract low dimensional abstract feature representations from high-dimensional raw RF ultrasound data. We study a number of methods following the successful ``Siamese neural networks" concept (see for instance ~\cite{chen2020simple,he2020momentum,zbontar2021barlow,bardes2021vicreg}), where two different views (augmentations) of an instance image are mapped to two low-dimensional representation vectors using the shared backbone neural network (Figure \ref{fig:method}.1).

Formally, given an extracted raw RF patch $x_i^{\text{raw}}$, two transformations $t$ and $t'$ are sampled from a distribution $\tau$, and two augmented views are produced, $x_i=t(x_i^{\text{raw}})$ and $x'_i=t'(x_i^{\text{raw}})$. The augmented data are then mapped to representation vectors $h_i$ and $h'_i$ using the backbone network:
\begin{equation}\label{eq:feature_extraction}
    h_i=f_\theta(x_i) .
\end{equation}
Next, the representations are projected to $z_i$ and $z'_i$ using the projection network $z_i=g_\phi(h_i)$. During self-supervised training, $\theta$ and $\phi$ are tuned to minimize a cost function with respect to the pairs $(z_i, z'_i)$.

We propose to use the recent Variance-Invariance-Covariance Regularization (VICReg)~\cite{bardes2021vicreg} method which we found to have the best performance among the methods tested. VICReg maximizes the agreement between representations $z_i$ and $z'_i$ using the mean-squared error loss. A trivial solution would be for the network to return constant output regardless of input, a phenomenon called representation collapse. 
VICReg avoids collapse by encouraging the variance across features in a batch to be above a certain threshold and the covariance between features to be as low as possible; this acts to maximize the information content of representations.


The VICReg loss $\ell(Z, Z')$ is the weighted sum of three regularization terms, called the invariance $s(Z, Z')$, variance $v(Z)$, and covariance $c(Z)$ losses defined as follows: 
\begin{align}\label{eqn:vicreg_loss}
    \begin{split}
    \ell(Z, Z') & = \lambda s(Z,Z') \\ 
                         & \quad \quad + \mu [v(Z)+v(Z')] \\
                           & \quad \quad \quad \quad+ \nu [c(Z)+c(Z')]   , 
    \end{split}
\end{align}
where the weights $\lambda$, $\mu$, and $\nu$ are hyperparameters and $Z$ and $Z'$ denote the batches of projection vectors $z_i$ and $z'_i$ (i.e. $Z$ is the matrix whose rows are $[z_1, z_2, ..., z_n]$ for batch size $n$). $z^{j}$ denotes the $j$'th column of Z, that is the vector composed of the $j$'th feature of each projection vector in the batch.


The invariance term $s(Z, Z')$ is the mean-squared error loss between each pair of $z_i$ and $z'_i$:
\begin{equation}
    s(Z, Z') = \frac{1}{n}{\sum^n_{i=1}}\|z_i-z'_i\|^2 .  
\end{equation}
Minimizing this term forces the network to learn features which are invariant under augmentations of the same input data.

The variance regularization term $v(Z)$ is defined as the hinge function of standard deviation of the projections along with batch dimension, namely:  
\begin{equation}
    v(Z) = \frac{1}{d}{\sum^d_{j=1}}\max(0, \gamma - \sigma(z^j)), 
\end{equation}
where $d$ is the feature dimension and 
\begin{equation}
    \sigma(z^j) = \sqrt{\epsilon + \text{Var}(\{z^j_i | i=1,..., n\})},
\end{equation}
 and $\gamma$ is a threshold value set to $1$ in our experiments, and  $\epsilon$ is a small number added to reduce numerical instability in the standard deviation. Minimizing this loss maintains variance in each feature of the representations, thereby avoiding mode collapse.

 The covariance regularization term $c(Z)$ is: 
\begin{equation}
    c(Z)=\frac{1}{d}{\sum_{i\neq{j}}}C(Z)_{ij}^2,
\end{equation}
where $C(Z)$ is the covariance matrix of $Z$ given by: 
\begin{equation}
    C(Z) = \frac{1}{n-1}{\sum^n_{i=1}}(z_i-\Bar{z})(z_i-\Bar{z})^T,  \Bar{z}=\frac{1}{n}{\sum^n_{i=1}z_i}. 
\end{equation}
Notice $c(Z)$ is the sum of the off-diagonal coefficients of covariance matrix $C$. By minimizing this, the covariance between features is forced to be close to 0 which minimizes redundancies due to intercorrelations between features.
 
During pretraining, the parameters $\theta$ and $\phi$ of the feature extractor network $f_\theta(.)$ and projection network $g_\phi(.)$ are optimized to minimize the VICReg loss across many mini-batches of augmented data pairs.



\subsection{Supervised Finetuning}

Followed self-supervised learning, the feature extractor $f_\theta(.)$ is paired with a linear classification head $k_\psi(.)$ which projects feature representations of input patches to a two-dimensional vector corresponding to the probability for the benign (0) and cancer (1) class. This network is trained to minimize the cross-entropy loss between the predicted probability and the ground truth class labels within a labeled dataset. Following common practice in SSL literature (e.g., see VICReg~\cite{bardes2021vicreg}), we consider either \emph{linear finetuning}, when only the linear layer weights $\psi$ are optimized, and \emph{semi-supervised finetuning}, where both the linear layer and feature extractor weights $\psi$ and $\theta$ are optimized. 

\begin{figure*}[h!]
    \centering
    \includegraphics[scale=.6]{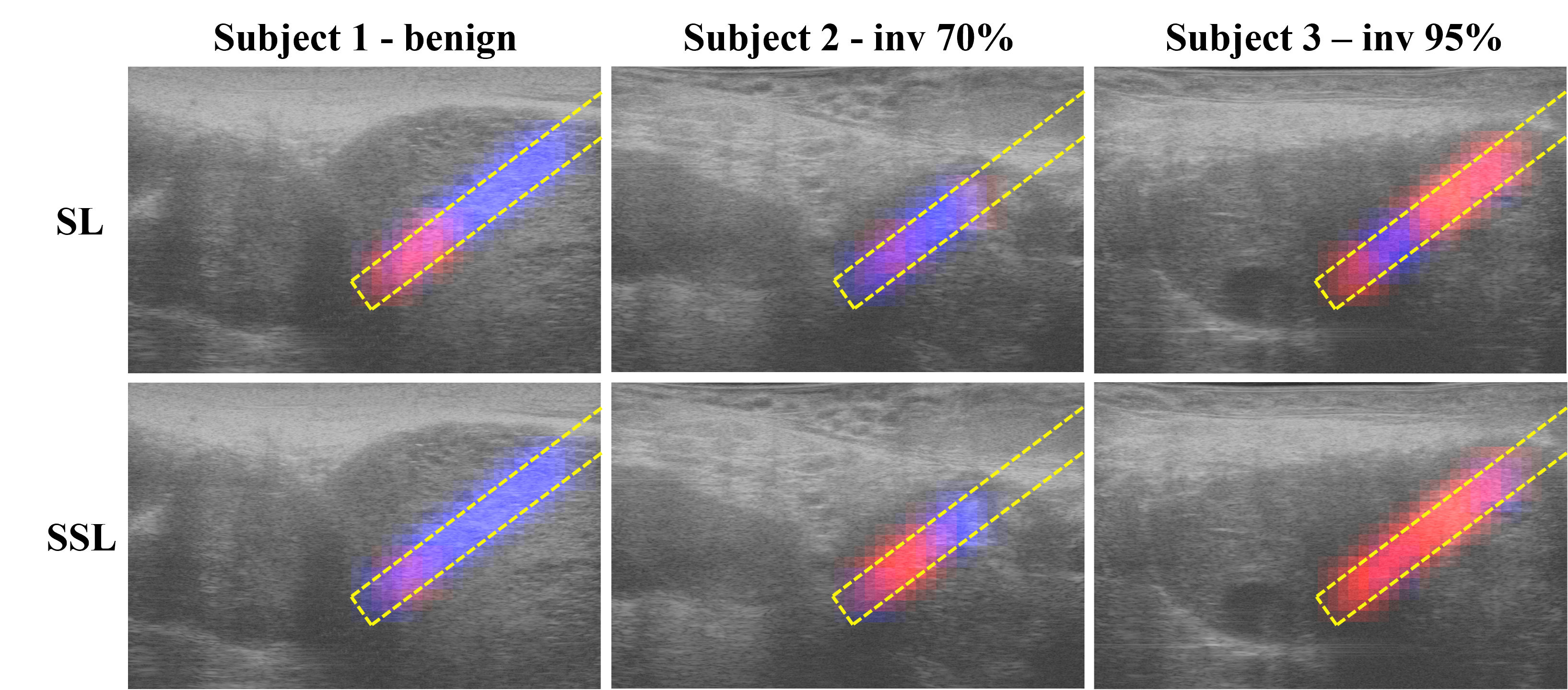}
    \caption{Demonstration of cancer detection models. For each ultrasound image, the model predicts whether patches in the intersection of prostate and needle trace region (yellow dashed line) are benign (blue) or malignant (red). Three subjects are shown demonstrating model performance on benign (Subject 1, left), malignant with 70\% involvement (Subject 2, middle) and malignant with 95\% involvement (Subject 3, right) tissue. We demonstrate two models, namely the fully supervised model (SL) and self-supervised model (SSL) when all layers are finetuned. Note that for all subjects the SSL model makes fewer incorrect prediction, and the ratio of malignant (red) to benign (blue) using the SSL model better correlates with the pathology reported involvement of cancer in each biopsy core.}
    \label{fig:heatmaps}
\end{figure*}

\begin{figure}
    \centering
    \includegraphics[scale=.7]{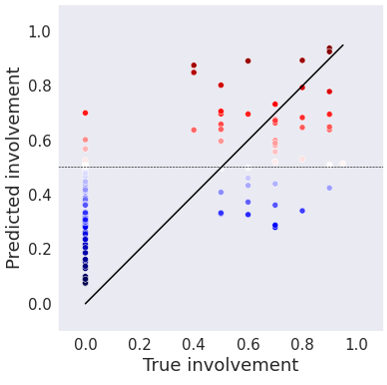}
    \caption{Scatter plot of predicted cancer involvement using the self-supervised model (SSL) versus true involved as reported by histopathology. Each point corresponds to a biopsy core in the UVA test data. Predicted involvement is calculated as the sum of predicted classes (benign or cancer) for patches within a core. In an ideal scenario, predicted involvement would match the true involvement and the points would lie on the diagonal line. The color range (blue to red) correlates with the predicted cancer involvement, with darker blue indicating low involvement and darker red indication indicating high involvement.}
    \label{fig:scatter}
\end{figure}
\subsection{Quantitative Evaluation}

We measure our model's performance in terms of cancer classification in two ways: First, we measure its performance for patch classification. As previously noted, patch-wise labels are weak labels and classification performance on these weak labels is not a perfect measurement of true performance. (To illustrate this, consider, a model \emph{correctly} predicting the label ``benign" on a benign patch from a cancerous core would be registered as a \emph{false} prediction, because that patch would be labeled according to the overall core label of ``cancer"). Still, we assume that patch classification performance on weak labels strongly correlates with true performance. Second, we measure the performance of the model for core classification by defining the predicted class probabilities for a core to be the mean of predicted classes (0 or 1) for patches within that core. For both patchwise and corewise performance we measure balanced accuracy (ACC-B, the average of sensitivity and specificity), average precision (Avg-Prec) and area under the receiver operating characteristic curve (AUROC).


When computing the metrics, we ignore cores with cancer involvements of less than $40\%$. This choice follows conventions used by the previous literature \cite{rohrbach2018high,shao2020improving,gilany2022towards} and is justified as (1) cores with low involvement cores are less likely to contain features representative of cancer and (2) weak labelling means that using a low involvement core will result in more incorrectly labeled patches than correctly labeled ones. Training with these cores is therefore problematic and performance metrics using them are invalid. 

\subsection{Qualitative Evaluation}
\label{subsection:qualitative}
To demonstrate the output of our models, we allow the models to predict the tissue type (benign or malignant) of each patch within the intersection of prostate region and needle. These are compiled into a heatmap, where regions of blue correspond to benign predictions and regions of red correspond to malignant predictions. The heatmaps are overlaid over the corresponding b-mode image to show the model's prediction of the spread of cancer. We can compare the output of model to the involvement of cancer estimated in the pathology reports. The number of ``cancer" predictions compared to total predictions can be considered the ``predicted involvement" of the model for that needle region, and can be approximated visually or calculated numerically. If the predicted involvement is close to true involvement, this reflects good model performance.


\begin{table*}[ht]
  \caption{Comparison of supervised $vs.$ self-supervised models on prostate cancer detection. The labeled dataset for both experiments was $\mathcal{D}^{\text{UVA}}_{\text{needle}}$.}
  \label{SL_SSL_compare}
  \centering
  \begin{tabular}{l c c c c c c c c}
    \toprule
    \textbf{Method} & Pretrain & Finetuning  & AUROC & Avg-Prec & ACC-B & Patch-AUROC & Patch-Avg-Prec & Patch-ACC-B\\
    
    \midrule
    
    \multirow{2}{10em}{Supervised} & None & - & 87.87\scriptsize{$\pm$2.09} &86.23\scriptsize{$\pm$2.79} & 76.17\scriptsize{$\pm$6.05}& 74.39\scriptsize{$\pm$1.59}& 71.03\scriptsize{$\pm$0.18} & 66.76\scriptsize{$\pm$1.41}\\
    
     & ImageNet & - & 87.83\scriptsize{$\pm$1.70} & 86.83\scriptsize{$\pm$1.60} & N/A& 73.83\scriptsize{$\pm$1.70} & 71.47\scriptsize{$\pm$1.90} & N/A\\
    \midrule
    
    \multirow{1}{10em}{EDL+Co-teaching \cite{gilany2022towards}} & None & - & 87.76\scriptsize{$\pm$1.82} &N/A & 77.79\scriptsize{$\pm$4.21}& N/A& N/A & 71.25\scriptsize{$\pm$1.16}\\
    
    \midrule
    
    \multirow{4}{*}{VICReg} & \multirow{2}{*}{$\mathcal{D}^{\text{UVA}}_{\text{needle}}$} & Linear & 89.83\scriptsize{$\pm$1.27} & 88.77\scriptsize{$\pm$1.74} & 80.73\scriptsize{$\pm$2.79} & 79.43\scriptsize{$\pm$0.85} & 78.40\scriptsize{$\pm$0.71} & 70.71\scriptsize{$\pm$1.24}\\
    
     &  & Semi-sup & 89.30\scriptsize{$\pm$3.01} & 87.59\scriptsize{$\pm$3.70} & 77.94\scriptsize{$\pm$9.67} & 78.23\scriptsize{$\pm$2.48}& 76.70\scriptsize{$\pm$2.62} & 68.31\scriptsize{$\pm$3.74}\\

     & \multirow{2}{*}{$\mathcal{D}^{\text{UVA}}_{\text{prostate}}$} & Linear & \textbf{90.99\scriptsize{$\pm$3.18}} & \textbf{90.74\scriptsize{$\pm$3.59}} & \textbf{81.66\scriptsize{$\pm$3.55}} &
     \textbf{79.90\scriptsize{$\pm$2.89}} & \textbf{78.78\scriptsize{$\pm$3.03}} & \textbf{71.44\scriptsize{$\pm$2.36}}\\
    
     &  & Semi-sup & 90.60\scriptsize{$\pm$2.40} & 89.41\scriptsize{$\pm$3.18} & 80.91\scriptsize{$\pm$4.86} & 79.49\scriptsize{$\pm$1.62} & 78.25\scriptsize{$\pm$1.79} & 70.79\scriptsize{$\pm$2.00}\\

    \bottomrule
  \end{tabular}
\end{table*}

\begin{table*}[th!]
  \caption{Comparison of transfer learning performance following either supervised pretraining, self-supervised pretraining (VICReg), or no pretraining (baseline). }
  \label{SL_SSL_transfer_compare}
  \centering
  \begin{tabular}{l c c c c c c c c}
    \toprule
    \textbf{Method} & Pretrain  & Train & Test & Finetuning  & AUROC & Avg-Prec & Patch-AUROC & Patch-Avg-Prec\\
    
    \midrule
    \multirow{1}{*}{Supervised} & \multirow{1}{*}{None (baseline)}& \multirow{1}{*}{$\mathcal{D}^{\text{CRCEO}}_{\text{needle}}$} & \multirow{1}{*}{$\mathcal{D}^{\text{CRCEO}}_{\text{needle}}$} & - & 67.73\scriptsize{$\pm$3.63} &  62.67\scriptsize{$\pm$3.50} &59.41\scriptsize{$\pm$1.33} &57.18\scriptsize{$\pm$1.88}\\

    \midrule  
    \multirow{2}{*}{Supervised} & \multirow{2}{*}{$\mathcal{D}^{\text{UVA}}_{\text{needle}}$}& \multirow{2}{*}{$\mathcal{D}^{\text{CRCEO}}_{\text{needle}}$} & \multirow{2}{*}{$\mathcal{D}^{\text{CRCEO}}_{\text{needle}}$} & Linear & 70.44\scriptsize{$\pm$5.10} &  64.95\scriptsize{$\pm$3.64} &62.33\scriptsize{$\pm$1.95} & 59.92\scriptsize{$\pm$1.67}\\

     &  &  & & Semi-sup & 71.84\scriptsize{$\pm$4.60} &  66.43\scriptsize{$\pm$4.95} & 62.67\scriptsize{$\pm$2.06} & 59.23\scriptsize{$\pm$2.18}\\

    \midrule

    \multirow{2}{*}{VICReg} & \multirow{2}{*}{$\mathcal{D}^{\text{UVA}}_{\text{prostate}}$}  & \multirow{2}{*}{$\mathcal{D}^{\text{CRCEO}}_{\text{needle}}$} & \multirow{2}{*}{$\mathcal{D}^{\text{CRCEO}}_{\text{needle}}$} & Linear & 70.35\scriptsize{$\pm$3.05} & 65.81\scriptsize{$\pm$2.60} & 65.64\scriptsize{$\pm$2.14} & 62.30\scriptsize{$\pm$2.34}\\

    &  & & & Semi-sup & \textbf{75.33\scriptsize{$\pm$5.59}} & \textbf{69.40\scriptsize{$\pm$4.30}} & \textbf{67.12\scriptsize{$\pm$3.24}} & \textbf{64.65\scriptsize{$\pm$3.33}}\\
    
    \bottomrule
  \end{tabular}
\end{table*}

\section{Experiments}

We designed our experiments to answer four key questions relevant to the clinical application of our model:

\begin{enumerate}
    \item Does using SSL improve the performance of models compared to SL alone?
    \item Does SSL outperform SL on  transfer learning for downstream tasks (e.g., PCa detection on a different dataset)?
    \item Are SSL models more robust to inter-center distribution shifts in PCa data than their SL counterparts? 

    \item Are there statistically significant performance differences between different SSL methods for PCa detection?
\end{enumerate}

To answer these questions, we designed four experiments:

\paragraph{Comparing SSL to supervision} 
We compare self-supervised pretraining followed by finetuning to fully supervised learning. For a stronger comparison, we consider several supervised baselines: these include random initialization, ImageNet pretraining, and training with vs. without augmentations. We also compare to the model of Gilany et al.~\cite{gilany2022towards} which uses supervision in conjunction with co-teaching to handle label noise and uncertainty (this model used the same testing set, making the comparison possible). 

\paragraph{Testing transfer learning} 
We test transfer learning performance by using the UVA dataset as a pretraining dataset and transferring models to the CRCEO dataset, allowing the model to re-tune its weights. We compare the success of supervised pretraining to self-supervised pretraining and no pretraining. 

\paragraph{Testing robustness to distribution shift}
To test robustness to distribution shift, we train the model on the UVA dataset and test the model on the CRCEO dataset. The difference between this and the transfer learning experiment is that in this case, the model does \emph{not} retune its weights. We compare the distribution shift performance of supervised and self-supervised models to the ``home" distribution performance of a supervised model trained from scratch on CRCEO.

\paragraph{Comparing SSL methods}
We compared the quality of representations learned using three different SSL methods, which may be considered as representative of three broad classes: SimCLR~\cite{chen2020simple} representing the contrastive learning family, BYOL~\cite{he2020momentum} representing the momentum-teacher family, and VICReg~\cite{bardes2021vicreg} representing the more recent non-contrastive family. We measure linear finetuning performance which is a direct measurement of the quality of SSL representations (only a linear layer is trained; the feature extractor weights learned using SSL are unchanged).

\emph{Implementation:} For SSL pre-training and linear evaluation, we use the Adam~\cite{kingma2014adam} optimizer. For fine-tuning and supervised learning, we use the NovoGrad~\cite{ginsburg2019stochastic} optimizer which we found resulted in improved training stability. For all training protocols, we use learning rate schedule consisting of linear warmup to 1e-4 over 10 epochs followed by a cosine annealing schedule over the remaining epochs. This schedule was chosen based on precedent in SSL literature (e.g., Data2Vec~\cite{baevski2022data2vec}, VICReg~\cite{bardes2021vicreg}), and the base rate of 1e-4 was selected empirically using a hyper-parameter search. Self-supervised pretraining was carried out for 200 epochs, while finetuning or fully supervised learning converged quickly and was only done for 50 epochs. For VICReg pretraining, we use variance, invariance and covariance loss weights of 25, 25, 1, respectively, following the original paper~\cite{bardes2021vicreg}. In all experiments, a random subset of the training set was used for cross-validation. 
Following training, we restore the weights from the epoch during which the best AUROC for the validation set was recorded, and measure the performance on the test set. Each experiment is run 16 times with different random model initializations and random train/validation splits; mean and standard deviation of performance across runs is reported.



\section{Results and Discussion}

\subsection{Self-Supervised Models Outperform Supervised Models} \label{exp:comparisonSSLSL}

\begin{table*}[t]
  \caption{Comparison of SL and SSL on distribution shift}
  \label{tab:dist_shift}
  \centering
  \begin{tabular}{l c c c c c c c c}
    \toprule
    \textbf{Method} & Pretrain  & Train & Test & Finetuning  & AUROC & Avg-Prec & Patch-AUROC & Patch-Avg-Prec\\
    
    \midrule
    
    \multirow{1}{*}{Supervised} & \multirow{1}{*}{None (baseline)}& \multirow{1}{*}{$\mathcal{D}^{\text{CRCEO}}_{\text{needle}}$} & \multirow{1}{*}{$\mathcal{D}^{\text{CRCEO}}_{\text{needle}}$} & - & 67.73\scriptsize{$\pm$3.63} &  62.67\scriptsize{$\pm$3.50} &59.41\scriptsize{$\pm$1.33} &57.18\scriptsize{$\pm$1.88}\\

    \midrule
    
    \multirow{1}{*}{Supervised} & None  & \multirow{1}{*}{$\mathcal{D}^\text{UVA}_\text{needle}$} & \multirow{1}{*}{$\mathcal{D}^{\text{CRCEO}}_{\text{needle}}$} & - & \textbf{74.96\scriptsize{$\pm$3.85}} &  \textbf{68.91\scriptsize{$\pm$3.55}} & 66.94\scriptsize{$\pm$1.39} & 64.42\scriptsize{$\pm$1.41}\\

    \midrule

    \multirow{2}{*}{VICReg} & \multirow{2}{*}{$\mathcal{D}^\text{UVA}_\text{prostate}$}  & \multirow{2}{*}{$\mathcal{D}^\text{UVA}_\text{needle}$} & \multirow{2}{*}{$\mathcal{D}^\text{CRCEO}_\text{needle}$} & Linear & 71.90\scriptsize{$\pm$2.10} & 65.14\scriptsize{$\pm$1.30} & 65.80\scriptsize{$\pm$1.60} & 64.04\scriptsize{$\pm$1.70}\\

    &  & & & Semi-sup & 72.71\scriptsize{$\pm$5.79} & 66.33\scriptsize{$\pm$4.68} & \textbf{69.14\scriptsize{$\pm$2.08}} & \textbf{67.33\scriptsize{$\pm$2.32}}\\
    
    \bottomrule
  \end{tabular}
\end{table*}

\begin{table*}[t]
  \caption{Comparison of linear evaluation performance for different SSL methods on ultrasound RF data. All methods are pretrained on $\mathcal{D}^{\text{UVA}}_{\text{prostate}}$ and further evaluated by training on $\mathcal{D}^{\text{UVA}}_{\text{needle}}$.}
  \label{tab:compare_diff_ssl_methods}
  \centering
  \begin{tabular}{l c c c c}
    \toprule
    \textbf{Method} 
                        &      AUROC    &  Avg-Prec           & Patch-AUROC    &  Patch-Avg-Prec\\ 
    \midrule
    \multirow{1}{10em}{VICReg} & \textbf{90.99\scriptsize{$\pm$3.2}} & \textbf{90.74\scriptsize{$\pm$3.6}} & \textbf{79.90\scriptsize{$\pm$2.9}} & \textbf{78.78\scriptsize{$\pm$3.0}}\\
    
    \multirow{1}{10em}{SimCLR} & 82.90\scriptsize{$\pm$}3.0 & 81.23\scriptsize{$\pm$}3.0 &
    73.52\scriptsize{$\pm$}3.1 &
    70.82\scriptsize{$\pm$}2.8\\
    
    \multirow{1}{10em}{BYOL} & 
    51.25\scriptsize{$\pm$}5.8& 
    52.31\scriptsize{$\pm$}3.8&
    50.75\scriptsize{$\pm$}2.8&
    51.49\scriptsize{$\pm$}2.4\\
    
    \bottomrule
  \end{tabular}
\end{table*}

Table \ref{SL_SSL_compare} summarizes the comparison of SSL to SL using the quantitative metrics. Rows 1 and 2 are SL models with either random initialization or ImageNet pretraining. Row 3 is the evidential deep learning (EDL) + co-teaching model of \cite{gilany2022towards}. The only significant difference between the SL models is that the EDL+co-teaching model had superior patch-wise performance. The next four rows are the VICReg methods with different training datasets and finetuning protocols. Overall, comparing the best SSL vs. SL models, we see an improvement in core-wise AUROC ($+3\%$), core-wise Avg-Prec ($+3.9\%$), ACC-B ($+3.87\%$), patch-wise AUROC ($+5.51\%)$ and patch-wise Avg-Prec ($+7.3\%$). The only metric that does not significantly improve is Patch-ACC-B, where the EDL+co-teaching and SSL models are comparable. Improvements moving from baseline SL to SSL are more pronounced when considering patch-wise vs. core-wise metrics, so the SSL model is likely better at localizing cancer precisely.

Linear finetuning of SSL models tends to outperform semi-supervised finetuning by a small degree, and these differences are statistically significant ($p \leq 0.05$) for patch-wise metrics trained on the smaller $\mathcal{D}^{\text{UVA}}_{\text{needle}}$ dataset; this is likely because semi-supervised finetuning allows the feature extractor weights to be re-tuned via supervision and therefore reintroduces the risk of label memorization. Including more unlabeled data ($\mathcal{D}^{\text{UVA}}_{\text{prostate}}$ vs. $\mathcal{D}^{\text{UVA}}_{\text{needle}}$) resulted in consistent performance improvements when using SSL. These differences were  statistically significant for patch-wise metrics when using semi-supervised finetuning. This suggests that performance may improve further if even more unlabeled data were added, strengthening the case for SSL as a method to address data scarcity. Further experiments with larger volumes of data would be needed to confirm this.


To qualitatively study the performance of these models, we selected three biopsy cores and computed the heatmaps (Figure \ref{fig:heatmaps}) as explained in Section \ref{subsection:qualitative}. These correspond to a benign, malignant  ($70\%$ involvement), and malignant ($95\%$) moving from left to right. We compare predictions of SL and SSL models. Across all three patients, we see that area of ``cancer" predictions compared to total needle region is closer to the true involvement for the SSL model, agreeing with the quantitative measurements of improved performance. We also display a scatterplot (Figure \ref{fig:scatter}) summarizing the correlation of predicted involvement to true involvement across all cores in the test set for the SSL model. We see that the correlation is good in general, but many benign cores still have positive predicted involvement, showing that the model still has room for improvement by reducing false positive predictions.

\subsection{Self-Supervised Models Have Strong Knowledge Transfer}

Table ~\ref{SL_SSL_transfer_compare} summarizes quantitative evaluation results of the transfer learning experiment. In the first row we have the results of the baseline supervised model (trained from scratch on CRCEO) for comparison. The second row shows the models which are pretrained using SL on UVA and finetuned on CRCEO. We see a considerable performance increase compared to no pretraining. The third row shows the models pretrained using SSL, which have better performance yet. Of the two SSL models, the one using semi-supervised finetuning on CRCEO is significantly better. In summary, self-supervised training improves performance compared to no pretraining ($+7.6\%$ core-wise AUROC, $+6.73\%$ core-wise Avg-Prec, $+7.7\%$ patch-wise AUROC, $+7.5\%$ patch-wise Avg-Prec) and compared to supervised pretraining ($+3.48\%$ core-wise AUROC, $+3.3\%$ core-wise Avg-Prec, $+4.45\%$ patch-wise AUROC, $+4.73\%$ patch-wise Avg-Prec). 

As pretraining of any kind (either SSL or SL) strongly benefited performance on the CRCEO set compared to training from scratch, this confirms that transfer learning using other micro-ultrasound datasets is very beneficial for PCa detection. SSL is better than SL for transfer learning, but only when the feature extractor backbone is finetuned on the new data (semi-supervised finetuning). We reason that self-supervised learning
improves transfer performance by giving good \emph{initializations} of the feature extractor weights rather than producing a feature extractor which transfers directly; at least some re-tuning of these weights is needed to optimize performance. 

    


    

    

\subsection{Both Supervised and Self-supervised models are robust to Distribution Shift in RF Ultrasound Data}



Table~\ref{tab:dist_shift} summarizes quantitative results for the distribution shift experiment. Row 1 shows results for supervised training on CRCEO as a baseline comparison. We would expect Row 1 to have the best performance as it trains on the CRCEO data, whereas the other models train on UVA data. Row 2 shows results for SL training on UVA and direct testing on CRCEO. Surprisingly, this row has considerably improved metrics compared to the row 1 -- the model is not just robust to distribution shift, but actually has improved performance on this distribution shifted data. Row 3 shows results for SSL training on UVA and testing on CRCEO, which is also an improvement compared to the first row, and the semi-supervised version in particular outperforms the SL model on patchwise but not core-wise metrics. 

The counterintuitive findings of this experiment suggest some combination of the following: (1) that the models are robust to distribution shift; (2) that distribution shifts between centers are minimal; (3) that the UVA training set has particularly favorable features for training models compared to the CRCEO training set. (2) can be ruled out, as the presence of significant inter-center distribution shift for these datasets has been established by Shao et al.~\cite{shao2020improving}. We can conclude that likely that a combination of (1) and (3) occured, and that these models models are robust to distribution shift with no clear difference between SL and SSL. Further experimentation is warranted to investigate the influence of (3).  


    
    

    
    


\subsection{Comparison of Contemporary Self-Supervised Learning Methods}



 Table~\ref{tab:compare_diff_ssl_methods} summarizes the comparison between the various SSL methods tested. We found that methods rank as follows: VICReg, SimCLR, BYOL. VICReg led by a significant margin in terms of performance, and required relatively little tuning of hyperparameters to achieve stable training and good results. On the other hand, SimCLR still performs fairly well, but required a large batch size. BYOL does not appear to work at all for RF data, although this finding is limited in that we did not perform an exhaustive hyperparameter search. 

We speculate that the improved performance of VICReg may be due to the non-contrastive nature of the algorithm. In SSL, contrastive methods (e.g. SimCLR \cite{chen2020simple}) explicitly push feature representations of different input patches apart, which may rely on the unrealistic assumption that no instances in the same batch represent the same tissue type (and so \emph{should} have similar representations). Non-contrastive methods may be better because they do not use this assumption, although the benefits observed may be due to another property of the VICReg algorithm.

\section{Conclusion}

We carried out a multi-center study on prostate cancer detection using self-supervised representation learning for RF micro-ultrasound data. We argued that self-supervised learning is a promising approach to address several characteristic challenges associated with this dataset, including weak labeling, high data heterogeneity, distribution shift and data scarcity. We showed strong empirical evidence that self-supervised learning is beneficial for training models that can detect PCa and whose knowledge can be effectively transferred between data centers. We showed that recently proposed self-supervised learning methods outperform older methods on our data, giving hope for continued improvements in the future. Future work should focus on two directions: First, to use larger quantities of available RF data with larger models which may further improve PCa detection, and second to apply other techniques for handling weak labeling and uncertainty in combination with SSL for a unified learning approach.


\section*{Acknowledgment}
This work was supported by the Natural Sciences and Engineering Research Council of Canada and the Canadian Institutes of Health Research.

\bibliographystyle{IEEEtran}

\bibliography{refs}

\end{document}